\documentclass[aps,reprint,twocolumn,amsmath,amssymb,superscriptaddress,
floatfix]{revtex4-2}

\usepackage{bm}%
\usepackage[colorlinks=true,linkcolor=blue]{hyperref}
\usepackage{bbm}
\usepackage{mathrsfs}
\usepackage{amsmath}
\usepackage{graphicx}
\usepackage{dcolumn}% Align table columns on decimal point
\usepackage{url}
\usepackage{braket}
\usepackage{afterpage}
\allowdisplaybreaks

\begin{document}

\preprint{APS/123-QED}

\title{Three-dimensional optical characterization of magnetostrictive deformation in magnomechanical systems}%

\author{Xiaomin Liu}
\affiliation{College of Physics and Electronic Engineering, Shanxi University, Taiyuan 030006, China}
\affiliation{State Key Laboratory of Quantum Optics Technologies and Devices, Shanxi University, Taiyuan 030006, China}

\author{Jing Zhang}
\email{zjj@sxu.edu.cn}
\affiliation{College of Physics and Electronic Engineering, Shanxi University, Taiyuan 030006, China}
\affiliation{State Key Laboratory of Quantum Optics Technologies and Devices, Shanxi University, Taiyuan 030006, China}
\affiliation{Collaborative Innovation Center of Extreme Optics, Shanxi University, Taiyuan 030006, China}

\author{Jie Li}
\email{jieli007@zju.edu.cn}
\affiliation{School of Physics, Zhejiang University, Hangzhou 310027, China}

\author{Rongguo Yang}
\email{yrg@sxu.edu.cn}
\affiliation{College of Physics and Electronic Engineering, Shanxi University, Taiyuan 030006, China}
\affiliation{State Key Laboratory of Quantum Optics Technologies and Devices, Shanxi University, Taiyuan 030006, China}
\affiliation{Collaborative Innovation Center of Extreme Optics, Shanxi University, Taiyuan 030006, China}

\author{Jiangrui Gao}
\affiliation{State Key Laboratory of Quantum Optics Technologies and Devices, Shanxi University, Taiyuan 030006, China}
\affiliation{Collaborative Innovation Center of Extreme Optics, Shanxi University, Taiyuan 030006, China}

\author{Tiancai Zhang}
\affiliation{State Key Laboratory of Quantum Optics Technologies and Devices, Shanxi University, Taiyuan 030006, China}
\affiliation{Collaborative Innovation Center of Extreme Optics, Shanxi University, Taiyuan 030006, China}


\date{\today}

\begin{abstract} 
Magnomechanical systems with YIG spheres have been proven to be an ideal system for studying magnomechanically induced transparency, dynamical backaction, and rich nonlinear effects, such as the magnon-phonon cross-Kerr effect. 
Accurate characterization of the magnetostriction induced deformation displacement is important as it can be used for, e.g., estimating the magnon excitation number and the strength of the dynamical backaction. Here we propose an optical approach for detecting the magnetostrictive deformation of a YIG sphere in three dimensions (3Ds) with high precision. It is based on the deformation induced spatial high-order modes of the scattered field, postselection, and balanced homodyne detection. With feasible parameters, we show that the measurement precision of the deformation in $x$, $y$, and $z$ directions can reach the picometer level. We further reveal the advantages of our scheme using a higher-order probe beam and balanced homodyne detection by means of quantum and classical Fisher information. The real-time and high-precision measurement of the YIG sphere's deformation in 3Ds can be used to determinate specific mechanical modes, characterize the magnomechanical dynamical backaction and the 3D cooling of the mechanical vibration, and thus finds a wide range of applications in magnomechanics. 
\end{abstract}

\maketitle

\section{Introduction}

Cavity magnomechanics (CMM) studies the interaction among microwave cavity photons, ferromagnetic magnons, and magnetostriction induced long-lived phonons~\cite{031201}. In a typical CMM system, an yttrium-iron-garnet (YIG) sphere is used benefitting from several advantages of the YIG material, such as high spin density and low damping rate. For a YIG sphere with the diameter of hundreds of $\mu$m, the mechanical vibration is of a relatively low frequency, typically in the MHz range~\cite{1501286,031053,123601,243601,NC5652}, which is much smaller than both the cavity and magnon resonance frequencies in the GHz range. This results in a dispersive-type magnomechanical interaction~\cite{031201}, which is a root cause for the experimental observation of magnomechanically induced transparency~\cite{1501286}, dynamical backaction~\cite{031053,123601}, magnonic frequency combs~\cite{243601}, and the polaromechanical strong coupling at the mechanical sideband~\cite{NC5652}, and also a prerequisite assumed in many quantum protocols, e.g., for preparing entangled states~\cite{203601,213604,085001}. In particular, under an appropriate driving field, the magnomechanical dynamical backaction can result in the cooling (amplification) of the mechanical motion, accompanied with a decreasing (an increasing) amplitude of the motion and thus energy. This, in the experiment, is usually verified by the measurement of a broadened (reduced) linewidth of the mechanical mode in the spectrum, as observed under a low drive power~\cite{031053} and under a high drive power with the presence of strong Kerr nonlinearities~\cite{123601}. 

Here, we provide an optical approach to characterizing the magnomechanical dynamical backaction via real-time monitoring of the YIG sphere's deformation displacement in three dimensions (3Ds). Specifically, it is based on the deformation induced spatial high-order modes in the scattered field of the probe beam, and three postselection processes are employed to extract specific modes that correspond to the deformation displacement in 3Ds. Compared to the conventional measurement of the mechanical linewidth (damping rate), our approach offers a direct characterization of the magnetostrictive deformation, and can serve as an effective means to study the magnon induced backaction on the mechanical vibration via optically reading out the amplitude of the mechanical motion with high precision. In contrast to the one-dimensional detection schemes~\cite{120801,033507,121106}, our scheme allows for a 3D characterization of the deformation displacement and can thus be used for the measurement of the 3D cooling of the mechanical vibration of the YIG sphere.

\section{Measurement principle}
\begin{figure}[htbp]
\centering\includegraphics[width=8.6cm]{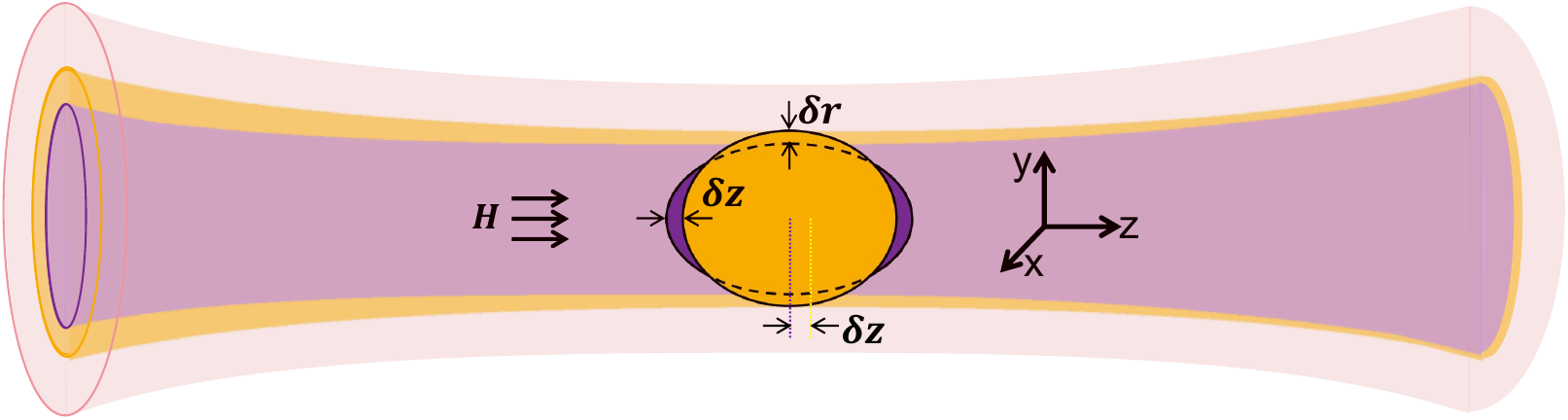}
 \caption{Schematic diagram of the magnetostrictive deformation of a YIG sphere. The YIG sphere is positioned in a uniform bias magnetic field $H$ and driven by a microwave field (not shown) whose magnetic component is perpendicular to the bias field. The magnetostrictive interaction leads to the geometric deformation of the YIG sphere, which, in each direction, can be decomposed into a small displacement fluctuation (orange ellipsoid) around an average displacement (purple ellipsoid). Here, the probe field is shown in pink and the scattered field corresponding to the deformed purple (orange) ellipsoid is shown in purple (orange). The small variation of the radial waist size and the shift of the axial waist position are denoted by $\delta r$ and $\delta z$, respectively. }
 \label{model}
\end{figure}

In a typical CMM system, the YIG sphere is placed inside a uniform bias magnetic field. When the YIG sphere is driven by another microwave field with its magnetic field perpendicular to the bias field, which can be provided by a microwave cavity or a loop antenna, all spins precess in phase forming a spin wave, i.e., a magnon mode. When the magnomechanical (magnetostrictive) interaction is further activated, the YIG sphere undergoes a deformation vibration, and the deformation displacement in each direction, $x$, $y$, or $z$, can be decomposed into a small displacement fluctuation around an average (constant) displacement, i.e., $q=\langle q \rangle + \delta q$~\cite{203601}, where $\langle q \rangle \gg \delta q$ is assumed, as depicted in Fig.~\ref{model}. Since the constant displacement $\langle q \rangle$ is usually much larger than $\delta q$ and it is determined by the average of the magnon excitation number $\langle m^\dagger m \rangle$, resulting from the dispersive magnomechanical interaction $\hbar g m^\dagger m q$~\cite{203601}, we are interested in and focus on the measurement of the small displacement fluctuation in 3Ds.

When a probe beam, whose waist is moderately larger than the size of the YIG sphere, is incident on a vibrational YIG sphere, the changes in the waist size and waist position of the scattered field correspond to the deformation of the sphere in the $xy$- and $z$ directions, respectively. In other words, one can obtain the 3D deformation information of the YIG sphere by measuring the corresponding spatial modes of the scattered field. It is well established that the scattering properties of homogeneous spherical particles across all size regimes—particularly those too large for Rayleigh theory and too small for geometric optics—can be described by Mie scattering theory~\cite{Wriedt2012,Wang2013}. Given that the diameter of commonly used YIG spheres is several hundreds of $\mu$m~\cite{031201}, the frequency of the incident beam is preferably selected in the terahertz range, where high-order modes of terahertz waves have already been experimentally generated and manipulated~\cite{1801328,20240711}.

In our scheme, Hermite-Gaussian (HG) modes are employed to measure the YIG deformation. The transverse distribution of a general HG mode along the $z$ axis can be expressed as
 \begin{equation}\begin{aligned}
 u_{nm}(x,y,z)=&\frac{1}{\sqrt{\pi 2^{n+m-1}n!m!}\omega(z)}H_n(\frac{\sqrt{2}x}{\omega(z)})H_m(\frac{\sqrt{2}y}{\omega(z)})\\& \times e^{-\frac{x^2+y^2}{\omega(z)^2}}e^{ik\frac{x^2+y^2}{2R(z)}}e^{-i(n+m+1)\phi_G(z)},
 \end{aligned}\end{equation}
where the beam radius at position $z$ is given by $\omega(z)=\omega_0\sqrt{1+(z/z_R)^2}$. Here, $\omega_0$ denotes the beam waist, and $z_R$ is the Rayleigh length, $z_R=\pi \omega_0^2/\lambda$ ($\lambda$ is the incident wavelength). $H_j$ ($j=n, m$) represents the $j$-th order Hermite polynomial, $k$ is the wave vector, the radius of curvature $R(z)=z+z_R^2/z$, and the Gouy phase $\phi_G(z)=\arctan(z/z_R)$.

\begin{figure}[b]
\centering\includegraphics[width=8.6cm]{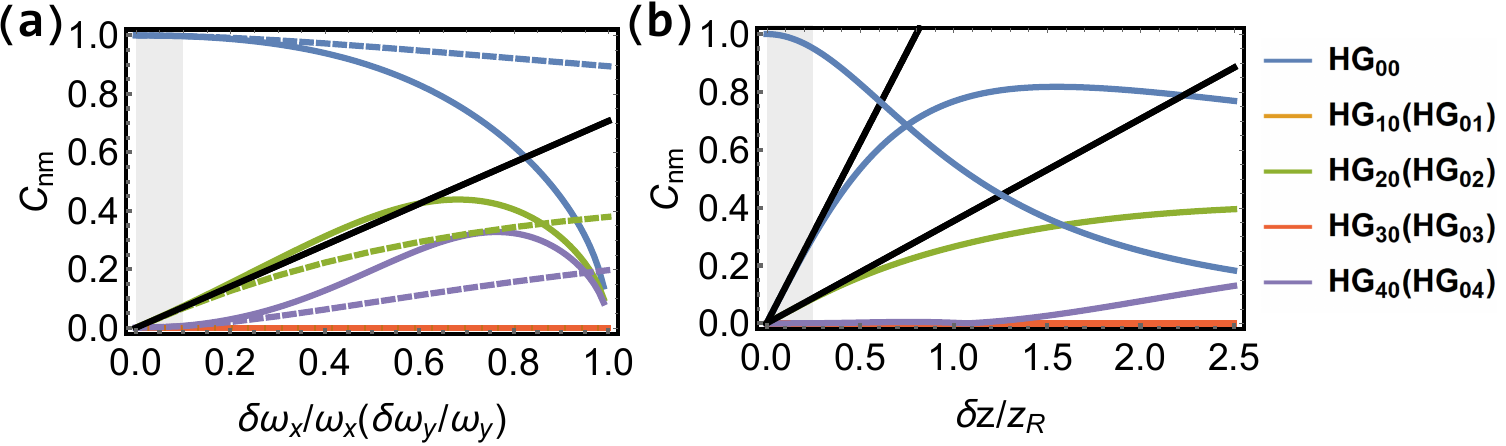}
\caption{Weight coefficients $C_{nm}$ of $\text{HG}_{nm}$ modes in the scattered field as a function of the variation in the waist radius (a) and the waist position (b) for a $u_{00}$ probe beam. The black lines denote the first-order approximations derived from Eq.~\eqref{eqqq2}. The solid (dashed) curves in (a) denote $\delta\omega_{x(y)}<0$ ($\delta\omega_{x(y)}>0$).}
\label{Cnm}
\end{figure}

When a Gaussian laser beam undergoes a variation in the waist size and waist position, specific high-order HG modes are induced~\cite{19945041,062001,2694}. 
To determine the dependence of the scattered field $\hat{E}_{s}$ on the deformation fluctuation, we take the first-order approximation of the Taylor expansion, given that the fluctuation in the scattered photon number induced by the deformation fluctuation is much smaller than the scattered photon number by the average deformation, i.e.,
\begin{equation}\label{eqqq2}
\begin{aligned}
 \hat{E}_{s}&\approx \hat{E}_{s}|_
 {\delta \omega_x=\delta \omega_y=\delta z=0}+\delta \omega_x \frac{\partial \hat{E}_{s}}{\partial\delta \omega_x}|_{\delta \omega_x=\delta \omega_y=\delta z=0}\\+&\delta \omega_y \frac{\partial \hat{E}_{s}}{\partial\delta \omega_y}|_{\delta \omega_x=\delta \omega_y=\delta z=0}+\delta z \frac{\partial \hat{E}_{s}}{\partial\delta z}|_{\delta \omega_x=\delta \omega_y=\delta z=0}.
\end{aligned}
\end{equation}
Here, $\delta \omega_x,\delta \omega_y$ are the variation of the waist radius in the $x$ and $y$ directions, respectively, and $\delta z$ is the shift of the waist position in the $z$ direction. For a $u_{00}$ probe beam, the waist variations in the $x$, $y$, and $z$ directions are related to the induced high-order fields $u_{20}$, $u_{02}$, and $(iu_{20}, iu_{02}, iu_{00})$, respectively ($\mathtt{Appendix~A}$).
The applicable range of the Taylor first-order approximation is denoted by the gray area in Fig.~\ref{Cnm}. Since the scattered fields $iu_{00}$ and $iu_{20} (iu_{02})$ induced by the shift $\delta z$ in the waist position have components of 93 \% and 7 \%, respectively~\cite{1964333,200700146849,1990}, it is a good approximation to consider only the $iu_{00}$ field to measure $\delta z$. In $\mathtt{Appendix~B}$, we provide more details on the induced second-order fields (simulated by COMSOL) and the far-field interference between the probe and the scattered fields.

\begin{figure}[t]
\centering\includegraphics[width=8.5cm]{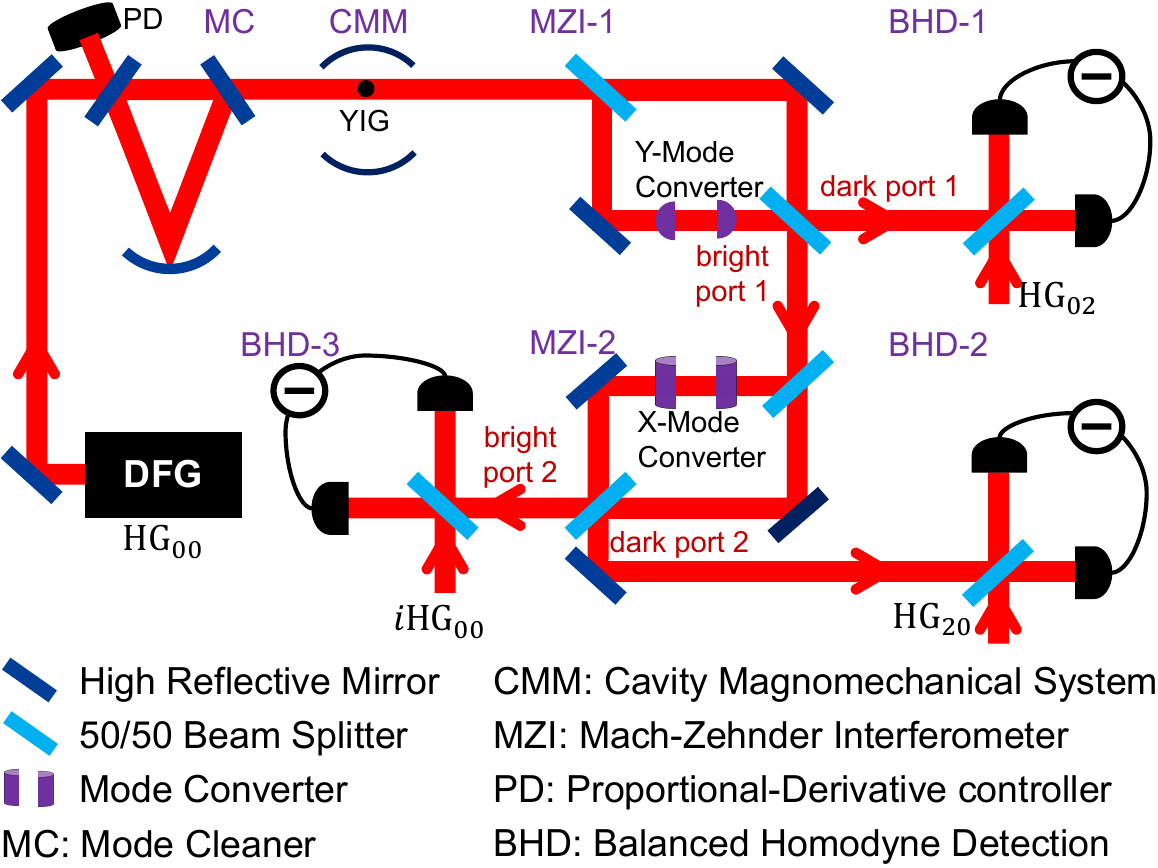}
\caption{Measurement scheme based on high-order modes of the scattered field, postselection, and balanced homodyne detection.}
\label{experiment}
\end{figure}

\section{Measurement scheme}

To simultaneously measure the small deformation displacement in the $x$, $y$, and $z$ directions with high accuracy, we adopt the weak measurement process, in which the postselection allows us to extract the target signals. The measurement scheme is depicted in Fig.~\ref{experiment}. A probe beam (the $\text{HG}_{00}$ mode) generated by the difference frequency generation (DFG) passes through a mode cleaner to achieve a purer $\text{HG}_{00}$ mode, which is then incident on the YIG sphere, e.g., placed inside a microwave cavity to form a CMM system. The resulting scattered field, which primarily contains the $u_{00}$, $iu_{00}$, $u_{20}$, and $u_{02}$ fields, sequentially enters two Mach-Zehnder interferometers (MZIs) denoted as MZI-1 and MZI-2. Constructive and destructive interferences occur for the $u_{02}$ and ($u_{00}$, $iu_{00}$, $u_{20}$) fields at the dark port 1, respectively, due to the Y-mode converter in MZI-1. Then, the deformation in the $y$ direction, i.e., $\delta\omega_y$, can be measured by the BHD-1 system with a local beam of the $\text{HG}_{02}$ mode. The output fields ($u_{00}$, $iu_{00}$, $u_{20}$) from the bright port 1 enter MZI-2 in which an X-mode converter is placed. In the same way, the information of the deformation in the $x$ direction $\delta\omega_x$ is included in the $u_{20}$ output field from the dark port 2, and can be measured by the BHD-2 system with a local beam of the $\text{HG}_{20}$ mode. In addition, the information of the deformation in the $z$ direction $\delta z$ is included in the $iu_{00}$ output field from the bright port 2, which can be measured with the BHD-3 system. Here, the X- and Y-mode converters, each consisting of two cylindrical lenses with a focal length of $f$, where $f(\omega_0)=\frac{\pi\omega_0^2}{\lambda}/(1+\frac{1}{\sqrt{2}})$, and a relative distance of $\sqrt{2}f$~\cite{200035}, can induce a phase flip for the $\text{HG}_{20}$ and $\text{HG}_{02}$ modes~\cite{102001}, respectively.

\begin{figure}[t]
\centering\includegraphics[width=8.5cm]{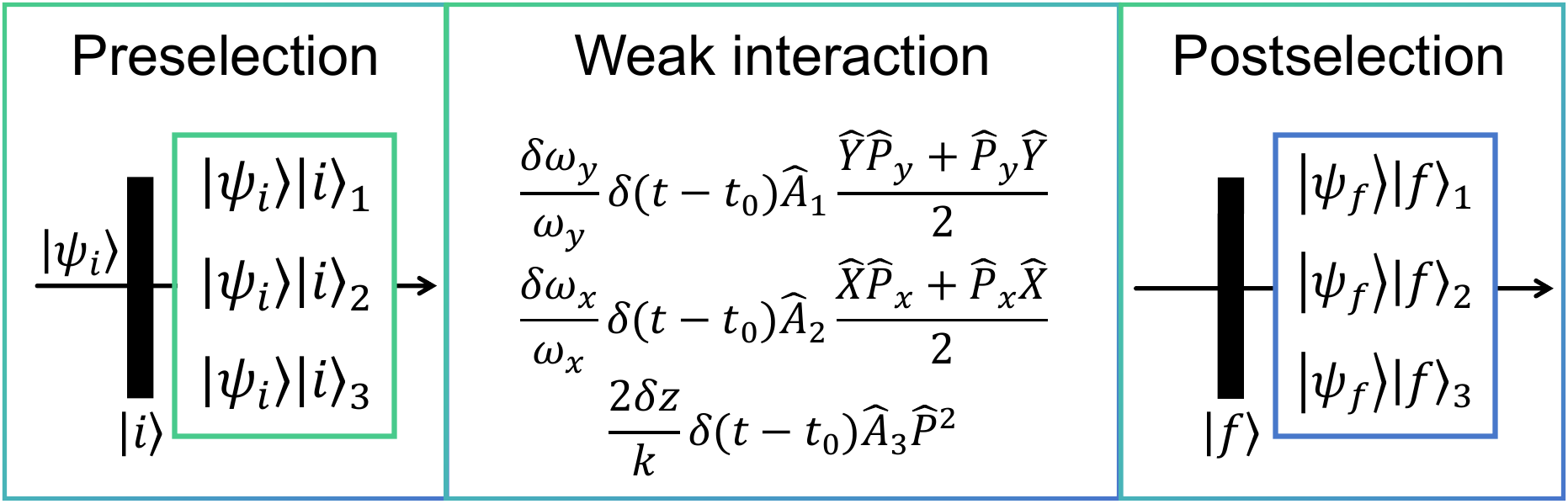}
\caption{Schematic diagram of the weak measurement process used in our scheme.}
\label{weak}
\end{figure}

A standard weak measurement process contains three stages: preselection, weak interaction, and postselection~\cite{1351}. To facilitate subsequent analysis of the quantum Fisher information in Sec.~\ref{Fisher}, the quantum formulation is employed. As illustrated in Fig.~\ref{weak}, the initial pointer state $|\psi_i\rangle$ evolves into the final state $|\psi_f\rangle$ after the weak interaction (scattering due to the small deformation fluctuations), while $|i\rangle$ and $|f\rangle$ denote the preselective and postselective system states. Generally, the interaction Hamiltonian can be defined as $\hat{H}_{\rm int}=\sum_i g_i\delta(t-t_0)\hat{A}_i\otimes\hat{\varepsilon}_i$, where $g_i$ denotes an element of the unknown parameter vector $\bm{g}=(g_1, g_2, \cdots, g_i, \cdots)$, $\hat{A}_i$ represents the measurement operator on the system, and $\hat{\varepsilon}_i$ stands for the general translation operator on the pointer. With this definition, the interaction Hamiltonian in our scheme can be written as ($\mathtt{Appendix~C}$)
\begin{equation}\label{HHHH}
\begin{aligned}
\hat{H}_{\rm int}=&\Big(\frac{\delta\omega_y}{\omega_y}\hat{A}_1\otimes\frac{\hat{Y}\hat{P}_y + \hat{P}_y\hat{Y}}{2}+\frac{\delta\omega_x}{\omega_x}\hat{A}_2\otimes\\&\frac{\hat{X}\hat{P}_x + \hat{P}_x\hat{X}}{2} +\delta z\hat{A}_3\otimes\frac{\hat{P}^2}{2k}\Big)\delta(t-t_0),
\end{aligned}
\end{equation}
where $\hat{X}$ ($\hat{P}_x$) and $\hat{Y}$ ($\hat{P}_y$) are the position (momentum) operators in the $x$ and $y$ directions, $\hat{P}^2=\hat{P}_x^2+\hat{P}_y^2$, and $\frac{\hat{X}\hat{P}_x + \hat{P}_x\hat{X}}{2}$ ($\frac{\hat{Y}\hat{P}_y + \hat{P}_y\hat{Y}}{2}$) and $\frac{\hat{P}^2}{2k}$ represent the scaling operation~\cite{Goldstein1980,Sakurai2020,Goodman1969} and the shearing transformation~\cite{1986lasers,2005quantum}, respectively. 
The final state $|\psi_f\rangle$ corresponds to the evolution of the initial state $|\psi_i\rangle$ under the evolution operator defined as $\hat{U}=e^{-i\int \hat{H}_{\rm int}dt}$, given by
\begin{equation}\label{psi}
\begin{aligned}
 |\psi_f\rangle&=(\langle f_1|i_1\rangle e^{-i\frac{A_{W1}\delta\omega_y}{\omega_y}\frac{\hat{Y}\hat{P}_y + \hat{P}_y\hat{Y}}{2}}+\langle f_2|i_2\rangle\times\\& e^{-i\frac{A_{W2}\delta\omega_x}{\omega_x}\frac{\hat{X}\hat{P}_x + \hat{P}_x\hat{X}}{2}}+\langle f_3|i_3\rangle e^{-i\frac{A_{W3}\delta z\hat{P}^2}{2k}})|\psi_i\rangle,
\end{aligned}
\end{equation}
where $|\psi_i\rangle=|\psi_{nm}\rangle$, $\langle q|\psi_{nm}\rangle=\psi_{nm}(x,y)=u_{nm}(x,y,0)$ describes the transverse field distribution at the beam waist, with $|q\rangle$ being the eigenstate of the position operator, and $\omega_x$ ($\omega_y$) denotes the beam waist radius corresponding to the average deformation displacement in the $x$ ($y$) direction. 
The weak value $A_{Wj}$ is defined as $A_{Wj}=\langle f_j|\hat{A}_j|i_j\rangle/\langle f_j|i_j\rangle$ ($j=1,2,3$). According to the corresponding recurrence relation of the Hermite polynomials ($\mathtt{Appendix~D}$) and considering only the first-order expansion and the deformation-induced Gouy phase shift, the final pointer state $|\psi_f\rangle$ for a probe beam of the fundamental $\text{HG}_{00}$ mode is
\begin{equation}\label{psif}\begin{aligned}
 |\psi_f\rangle
 =&\sum_{j=1}^3\langle f_j|i_j\rangle|\psi_{00}\rangle-\langle f_3|i_3\rangle\Big(\frac{iA_{W_3}}{2k\omega_{x}^2}+\frac{iA_{W_3}}{2k\omega_{y}^2}\\&+\frac{iA_{W_3}}{k\omega_{x}^2}+\frac{iA_{W_3}}{k\omega_{y}^2}\Big)\delta z|\psi_{00}\rangle\\&+\left(\langle f_2|i_2\rangle\frac{A_{W_2}}{2\omega_x}\delta\omega_x+\langle f_3|i_3\rangle\frac{iA_{W_3}}{2k\omega_{x}^2}\delta z \right)\sqrt{2}|\psi_{20}\rangle\\&+\left(\langle f_1|i_1\rangle\frac{A_{W_1}}{2\omega_y}\delta\omega_y+\langle f_3|i_3\rangle\frac{iA_{W_3}}{2k\omega_{y}^2}\delta z \right)\sqrt{2}|\psi_{02}\rangle\\
\approx&\sum_{j=1}^3\langle f_j|i_j\rangle|\psi_{00}\rangle-\langle f_3|i_3\rangle\Big(\frac{iA_{W_3}}{2k\omega_{x}^2}+\frac{iA_{W_3}}{2k\omega_{y}^2}\\&+\frac{iA_{W_3}}{k\omega_{x}^2}+\frac{iA_{W_3}}{k\omega_{y}^2}\Big)\delta z|\psi_{00}\rangle\\&+\langle f_2|i_2\rangle\frac{A_{W_2}}{\sqrt{2}\omega_x}\delta\omega_x|\psi_{20}\rangle+\langle f_1|i_1\rangle\frac{A_{W_1}}{\sqrt{2}\omega_y}\delta\omega_y|\psi_{02}\rangle,
\end{aligned}\end{equation}
which shows that, in consistency with the results derived from the scattered field, the final state $|\psi_f\rangle$ comprises of the components $|\psi_{20}\rangle$, $|\psi_{02}\rangle$ and $(i|\psi_{20}\rangle, i|\psi_{02}\rangle, i|\psi_{00}\rangle)$, corresponding to the deformations $\delta\omega_x$, $\delta\omega_y$ and $\delta z$, respectively. Here $\frac{\delta z}{k\omega_{x}^2}+\frac{\delta z}{k\omega_{y}^2}$ represents the deformation-induced Gouy phase shift. It is obvious that the weights of $i|\psi_{20}\rangle$ and $i|\psi_{02}\rangle$ are much smaller compared with that of $i|\psi_{00}\rangle$; therefore, one can approximately consider that $\delta z$ is determined only by $i|\psi_{00}\rangle$. This can also be understood from a classical perspective; see $\mathtt{Appendix~A}$ for detailed analyses.  The expression of the final state for a general HG probe beam is provided in $\mathtt{Appendix~D}$.

For deformations $\delta\omega_y$, $\delta\omega_x$ and $\delta z$, the corresponding preselection and postselection system states are ($\ket{i_1}=(\ket{\circlearrowleft}_1+\ket{\circlearrowright}_1)/\sqrt{2}$, $\ket{f_1}=(e^{i\theta/2}\ket{\circlearrowleft}_1-e^{-i\theta/2}\ket{\circlearrowright}_1)/\sqrt{2}$), ($\ket{i_2}=(\ket{\circlearrowleft}_1+\ket{\circlearrowright}_1)(\ket{\circlearrowleft}_2+\ket{\circlearrowright}_2)/2$, $\ket{f_2}=(e^{i\theta/2}\ket{\circlearrowleft}_1+ e^{-i\theta/2}\ket{\circlearrowright}_1)(e^{i\phi/2}\ket{\circlearrowleft}_2 -e^{-i\phi/2}\ket{\circlearrowright}_2)/2$) and ($\ket{i_3}=(\ket{\circlearrowleft}_1+\ket{\circlearrowright}_1)(\ket{\circlearrowleft}_2+\ket{\circlearrowright}_2)/2$, $|f_3\rangle=(e^{i\theta/2}\ket{\circlearrowleft}_1+e^{-i\theta/2}\ket{\circlearrowright}_1)(e^{i\phi/2}\ket{\circlearrowleft}_2+e^{-i\phi/2}\ket{\circlearrowright}_2)/2$), while the corresponding measurement operators are $\hat{A}_1=(\ket{\circlearrowleft}\bra{\circlearrowleft}_1-\ket{\circlearrowright}\bra{\circlearrowright}_1)$, $\hat{A}_2=(\ket{\circlearrowleft}\bra{\circlearrowleft}_2-\ket{\circlearrowright}\bra{\circlearrowright}_2)(\ket{\circlearrowleft}\bra{\circlearrowleft}_1+\ket{\circlearrowright}\bra{\circlearrowright}_1)$ and $\hat{A}_3=(\ket{\circlearrowleft}\bra{\circlearrowleft}_2+\ket{\circlearrowright}\bra{\circlearrowright}_2)(\ket{\circlearrowleft}\bra{\circlearrowleft}_1+\ket{\circlearrowright}\bra{\circlearrowright}_1)$, respectively. Here, $\ket{\circlearrowleft}_j$ and $\ket{\circlearrowright}_j$ ($j=1,2$) represent counterclockwise and clockwise paths of the MZI-$j$, respectively, and $\theta$ and $\phi$ denote the phase differences between the two paths of the MZI-1 and MZI-2, respectively. Thus, the final states at the dark port 1, dark port 2, and bright port 2 are obtained
\begin{equation}\label{bbbbb}
\begin{aligned}
|\psi_{f}^{d1}\rangle&=i\sin\frac{\theta}{2}e^{\frac{\delta\omega_y}{\omega_y}\frac{\hat{Y}\hat{P}_y + \hat{P}_y\hat{Y}}{2} \cot\frac{\theta}{2}}|\psi_{00}\rangle, \\|\psi_{f}^{d2}\rangle&=i\sin\frac{\phi}{2}\cos\frac{\theta}{2}e^{\frac{\delta\omega_x}{\omega_x}\frac{\hat{X}\hat{P}_x + \hat{P}_x\hat{X}}{2} \cot\frac{\phi}{2}}|\psi_{00}\rangle, \\|\psi_{f}^{b2}\rangle&=\cos\frac{\theta}{2}\cos\frac{\phi}{2}e^{-i\frac{\delta z}{z_R}-i\delta z\frac{\hat{P}^2}{2k}}|\psi_{00}\rangle,
\end{aligned}
\end{equation}
which indicate that the deformation in 3Ds can be achieved by measuring the output state from the corresponding port with a BHD system. Here $e^{-i\frac{\delta z}{z_R}}$ is the evolution corresponding to Gouy phase shift. It is worth mentioning that the above measurement scheme can also be analyzed from the perspective of the evolution of the probe field, as shown in $\mathtt{Appendix~A}$. 

The photon number differences of the three BHD systems are~\cite{6495}: $N_{1}^-=\sqrt{N_{\text{LO1}}}(\frac{\delta\omega_yA_{W1}\sqrt{NP_{s1}}}{\sqrt{2}\omega_y}+\frac{\delta \hat{X}_{02}}{2})$, $N_{2}^-=\sqrt{N_{\text{LO2}}}(\frac{\delta\omega_xA_{W2}\sqrt{NP_{s2}}}{\sqrt{2}\omega_x}+\frac{\delta \hat{X}_{20}}{2})$, and $N_{3}^-=\sqrt{N_{\text{LO3}}}(\frac{5\delta zA_{W3}\sqrt{NP_{s3}}}{4z_R}+\frac{\delta \hat{Y}_{00}}{2})$. Here, $N_{{\rm LO}j}=\frac{1}{\tau_r}\frac{P_{{\rm LO}j}\lambda_{{\rm LO}j}}{\hbar c}$ ($j=1,2,3$) is the photon number coming from the local beam to the detectors of the BHD-$j$ system during the time $1/\tau_r$, where $\tau_r$ is the resolution bandwidth of the detectors, $P_{{\rm LO}j}$ and $\lambda_{{\rm LO}j}$ denote the power and wavelength of the local beam, $\hbar$ is the reduced Planck constant, and $c$ is the speed of light. $P_{sj}=|\langle f_j|i_j\rangle|^2=P_{{\rm out}j}/P_{{\rm in}}^s$ is the corresponding postselection probability, with $P_{{\rm in}}^s=\frac{P_{{\rm in}}\int_{-\omega_x}^{\omega_x}\int_{-\omega_y}^{\omega_y}u_{00}dxdy}{\iint_{-\infty}^{\infty}u_{00}dxdy }$ being the power of the scattered beam, $P_{{\rm in}}$ being the probe power, and $P_{{\rm out}j}$ being the corresponding output power detected by the BHD-$j$ system. $NP_{sj}=\frac{1}{\tau_r}\frac{P_{{\rm out}j}\lambda}{\hbar c}$ corresponds to the output photon number incident on the detectors of the BHD-$j$ system, with $\lambda$ being the wavelength of the probe beam. Additionally, $\delta \hat{X}_{nm}=\delta \hat{a}_{nm}+\delta \hat{a}_{nm}^\dagger$ and $\delta \hat{Y}_{nm}=i(\delta \hat{a}_{nm}^\dagger-\delta \hat{a}_{nm})$ ($nm=00,02,20$) are the corresponding amplitude and phase fluctuation operators, with $\hat{a}_{nm}$ ($\hat{a}_{nm}^{\dagger}$) being the annihilation (creation) operator of the $\text{HG}_{nm}$ mode.

The corresponding signal-to-noise ratios (SNRs) are defined as: $R_1=(\frac{\sqrt{2}\delta\omega_y A_{W1}\sqrt{NP_{s1}}}{\omega_y}/\delta \hat{X}_{02})^2$, $R_2=(\frac{\sqrt{2}\delta\omega_xA_{W2}\sqrt{NP_{s2}}}{\omega_x}/\delta \hat{X}_{20})^2$, and $R_3=(\frac{5\delta zA_{W3}\sqrt{NP_{s3}}}{2z_R}/\delta \hat{Y}_{00})^2$. For a coherent probe beam with $\delta^2\hat{X}_{nm}=\delta^2\hat{Y}_{nm}=1$, and setting the SNR to 1, we obtain the corresponding minimum measurable deformation (MMD) in the three directions, i.e., $\delta\omega_{y{\rm min}}=\frac{\omega_y/A_{W1}}{\sqrt{2NP_{s1}}}$, $\delta\omega_{x{\rm min}}=\frac{\omega_x/A_{W2}}{\sqrt{2NP_{s2}}}$, and $\delta z_{\rm min}=\frac{2z_R/A_{W3}}{5\sqrt{NP_{s3}}}$. We note that the measurement precision can be further improved by using a higher-order probe beam $\text{HG}_{nm}$ with $n,m>0$, which yields a sensing gain of $n^2+n+1 $ ($m^2+m+1$) in the $x$ ($y$) direction~\cite{062001,2694}. The corresponding MMDs are modified as $\delta\omega_{y{\rm min}}^{m}=\frac{\omega_y/A_{W1}}{\sqrt{2NP_{s1}}\sqrt{(m^2+m+1)}}$, $\delta\omega_{x{\rm min}}^{n}=\frac{\omega_x/A_{W2}}{\sqrt{2NP_{s2}}\sqrt{(n^2+n+1)}}$, and $\delta z_{\rm min}^{m}=\frac{2z_R/A_{W3}}{5\sqrt{NP_{s3}}\sqrt{(\nu^2+\nu+1)}}$, where $\nu=\text{max}[m,n]$.

\begin{figure}[t]
\centering\includegraphics[width=8.6cm]{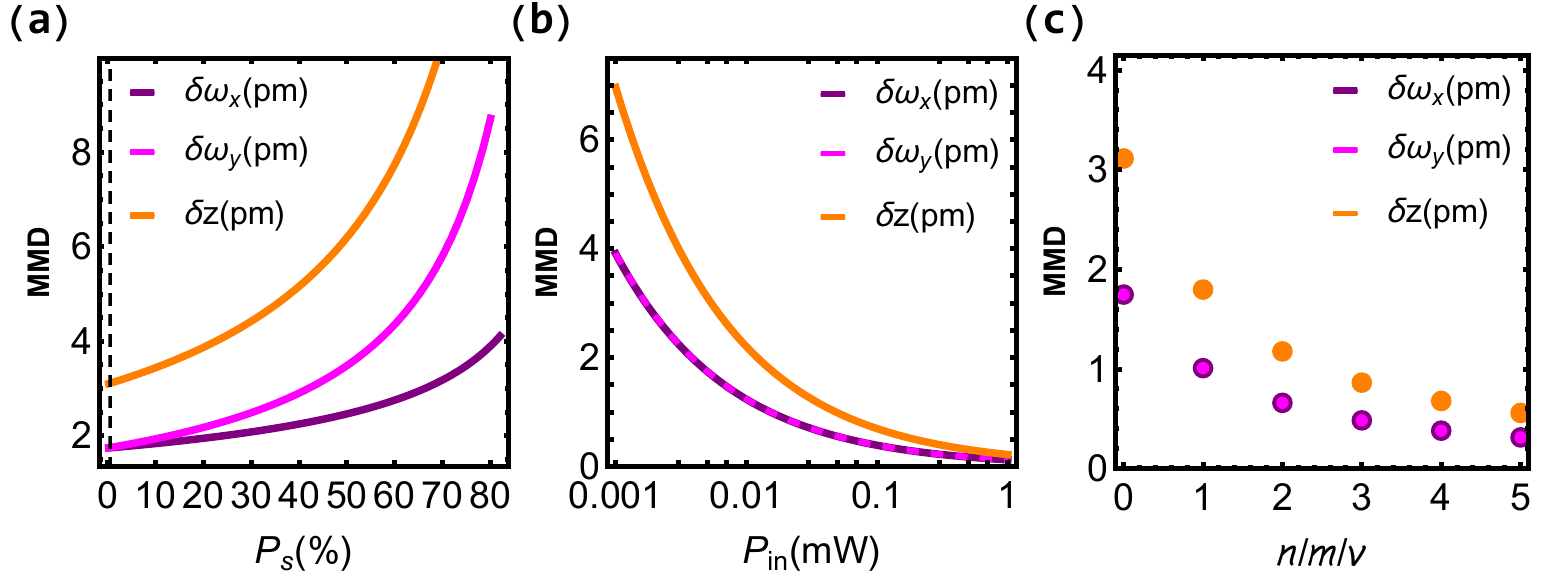}
\caption{Minimum measurable deformations (MMDs) versus (a) the postselection probability, (b) the probe power, (c) and the mode order of the probe beam. The purple, magenta, and orange dots (c) correspond to $\delta\omega_x$ versus $n$; $\delta\omega_y$ versus $m$; and $\delta z$ versus $\nu=\text{max}[m,n]$, respectively. We use a YIG sphere with the diameter of 250 $\mu$m, the detector resolution bandwidth $\tau_r=1$ Hz, the probe beam wavelength $\lambda=125$ $\mu$m, and the beam waist of the probe beam $\omega_0=150$ $\mu$m.}
\label{MMD}
\end{figure}

The effects of the postselection probability, the probe power, and the mode order of the probe beam on the MMDs are shown in Figs.~\ref{MMD}(a), \ref{MMD}(b) and \ref{MMD}(c), respectively. It is clear from Fig.~\ref{MMD}(a) that the MMDs decrease as the postselection probability reduces. Using moderate parameters in the experiment, e.g., a probe power of 5~$\mu\text{W}$ and a postselection probability of $0.5~\%$, the measurement precision of the deformation in the $x$, $y$, and $z$ directions can achieve 1.75, 1.75, and 3.11 pm, respectively. This picometer-level precision is sufficiently high for measuring the magnetostriction induced deformation of a YIG sphere~\cite{121106}. Further increasing the probe power can have a higher measurement precision, as shown in Fig.~\ref{MMD}(b). However, a too strong power should be avoided as the radiation pressure may induce a nonnegligible displacement. Figure~\ref{MMD}(c) indicates that higher-order probe beams can be adopted to improve the measurement precision. It is worth noting that for the probe beam being the $\text{HG}_{nm}$ mode, the deformation information is contained in the $\text{HG}_{n+2,m}$, $\text{HG}_{n,m+2}$, and $i\text{HG}_{nm}$ modes of the scattered field.

\section{Performance analysis based on Fisher information}
\label{Fisher}

The quantum Fisher information (QFI) is a central quantity for quantifying the sensitivity of a quantum state with respect to some parameter. A larger QFI indicates more information about the parameter to be estimated, corresponding to a higher measurement precision~\cite{1021}. In contrast, the classical Fisher information (CFI) is used to quantify the capability of the measurement scheme to extract the parameter information from the system, and a larger CFI indicates a stronger such capability of the scheme. In what follows, we adopt the tools of QFI and CFI to reveal the advantages of our scheme using a higher-order mode as the probe beam and the BHD system compared to other methods.
%\subsection{Quantum Fisher information}

\begin{figure}[t]
\centering\includegraphics[width=8.6cm]{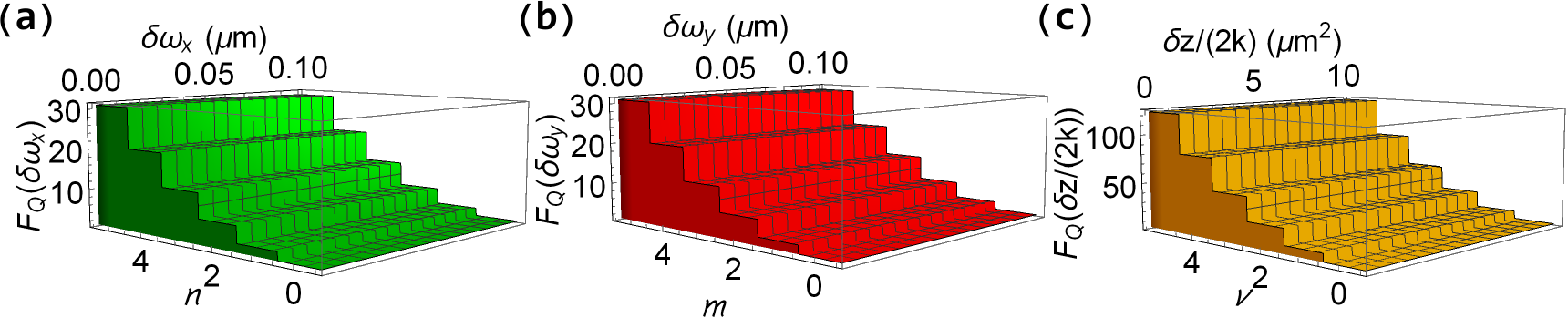}
\caption{Quantum Fisher information versus the mode order of the probe beam and the deformation in the (a) $x$, (b) $y$, and (c) $z$ direction.}
\label{quantum}
\end{figure}

For measuring the deformation of the YIG sphere in 3Ds, the matrix element of the QFI is defined as~\cite{034023,1021}
\begin{equation}
\begin{aligned}
{[F_{Q}]}_{ij}&=\text{Tr}\left[\rho_{\bm{g}}\cdot\frac{\hat{L}_i\hat{L}_j+\hat{L}_j\hat{L}_i}{2} \right] \\&=4\text{Re}\big(A_{Wi}^*A_{Wj} \langle\psi_i|\hat{\varepsilon}_i\hat{\varepsilon}_j|\psi_i\rangle \big),
\end{aligned}
\end{equation}
where $i,j=1,2,3$, and $\rho_{\bm{g}} = |\psi_f \rangle \langle \psi_f |$ denotes the density matrix of the measured state $|\psi_f \rangle$. $\hat{L}$ is the symmetric logarithmic derivative of the measured deformation parameter $\bm{g}=(g_1,g_2,g_3)=(\delta\omega_y,\delta\omega_x,\delta z)$ (with $g_1,g_2,g_3\ll 1$), which for a pure state $|\psi_f \rangle$ is defined as $\hat{L}_i = 2 \left( |\partial_i \psi_f \rangle \langle \psi_f | + |\psi_f \rangle \langle \partial_i \psi_f | \right)$, with $\partial_i =\partial/\partial_{g_i}$. The general translation operators are given by $\hat{\varepsilon}_1=(\hat{Y}\hat{P}_y + \hat{P}_y\hat{Y})/(2\omega_y)$, $\hat{\varepsilon}_2=(\hat{X}\hat{P}_x + \hat{P}_x\hat{X})/(2\omega_x)$, and $\hat{\varepsilon}_3=\hat{P}^2/(2k)$. We assume $\hat{A}_i=\hat{A}$, $A_{Wi}=A_W$, and $\omega_x=\omega_y=\sigma$. 
Considering the postselection process, the QFI is obtained as
\begin{equation}
\begin{aligned}
 F_Q=&|A_W|^2|\langle f|i\rangle|^2N \times\\&\text{diag}\left(\frac{n^2+n+1}{\sigma^{2}}, \frac{m^2+m+1}{\sigma^{2}},\frac{4(\nu^2+\nu+1)}{(2k)^2\sigma^4} \right).
\end{aligned}
\end{equation}
Under a low phase difference $\phi=\theta=0.01~\text{rad}$ and $\sigma=0.1$~mm, the results of the QFI are presented in Fig.~\ref{quantum}. Clearly, the QFI for measuring the deformation in 3Ds increases as the mode order of the probe beam increases, suggesting that a higher-order probe beam can achieve a higher measurement precision.

%\subsection{Classical fisher information}

On the other hand, the matrix element of the CFI is defined as~\cite{034023,043508}
\begin{equation}
\begin{aligned}
{[F_C]}_{ij}=\sum_{\lambda} \frac{1}{\langle \hat{\varepsilon}_{\lambda} \rangle} \left(\frac{\partial \langle \hat{\varepsilon}_{\lambda} \rangle}{\partial g_i}\frac{\partial \langle \hat{\varepsilon}_{\lambda} \rangle}{\partial g_j} \right),
\end{aligned}
\end{equation}
where $i,j=1,2,3$, and the measurement probability of the state $|\psi_f\rangle$ under the operator $\hat{\varepsilon}_{\lambda}$ is given by $\langle \hat{\varepsilon}_{\lambda} \rangle = \langle \psi_f | \hat{\varepsilon}_{\lambda} | \psi_f \rangle$, with $\hat{\varepsilon}_{\lambda}=|q\rangle \langle q|$. 
In the following, we shall use the CFI to compare our BHD method with the array detection (AD) method~\cite{1494855,013820}, which adopts $\pi$-phase flipping between specific regions of the transverse distribution of the probe field to match specific high-order modes of the scattered field. Notably, for detecting the first-order HG modes, the AD reduces to the split detection~\cite{053823,054068}, which is commonly used in the weak measurement experiments.
%\subsubsection{AD detection}
The CFI of measuring the deformation in 3Ds using the AD method and a $\text{HG}_{00}$ probe beam in the postselection scheme is obtained: 
\begin{equation} \begin{aligned}
 F_C^{\rm (AD)}&=|A_W|^2|\langle f|i\rangle|^2N \times\\&\text{diag}\left(\sqrt{\frac{2\sqrt{2/e}}{\pi}}\sigma^{-2}, \sqrt{\frac{2\sqrt{2/e}}{\pi}}\sigma^{-2}, \frac{1}{k^2}\sigma^{-4} \right).
 \end{aligned} \end{equation}
The factor $\sqrt{\frac{2\sqrt{2/e}}{\pi}}$ arises from the incomplete overlap between the induced $u_{20}$ and $u_{02}$ modes and the corresponding flipped modes~\cite{393}, due to a $\pi$ phase flip in the horizontal or vertical direction from $-\omega_0/2$ to $\omega_0/2$, i.e., $\int_{-\infty}^{-\frac{\omega_0}{2}}dr u_{00}u_{20}+\int^\infty_{\frac{\omega_0}{2}}dr u_{00}u_{20}-\int_{-\frac{\omega_0}{2}}^{\frac{\omega_0}{2}}dr u_{00}u_{20}=\sqrt{\frac{2\sqrt{2/e}}{\pi}}$ ($r=x$ or $y$).

\begin{figure}[t]
\centering\includegraphics[width=8.6cm]{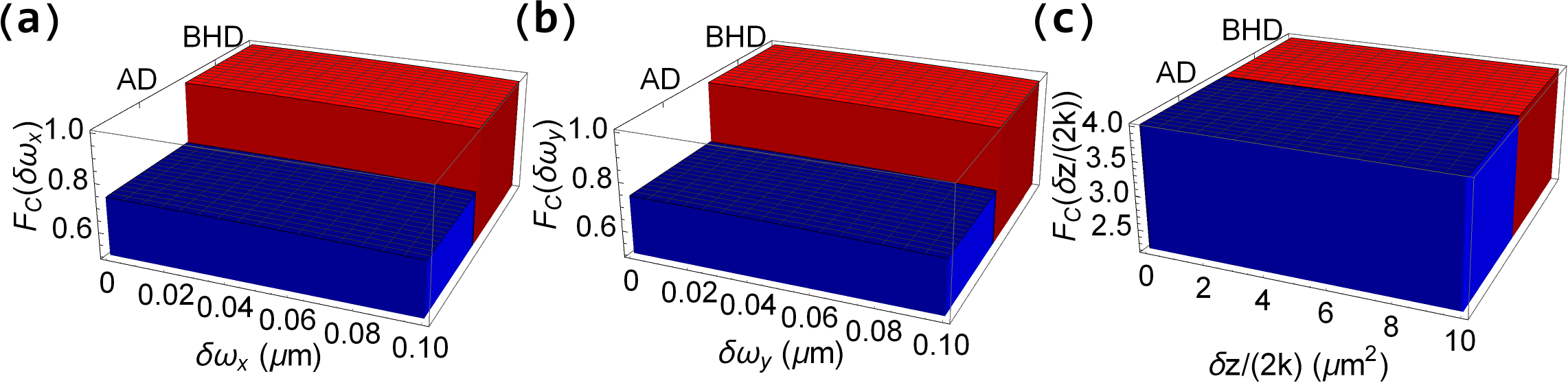}
\caption{Classical Fisher information of AD and BHD versus deformation in the (a) $x$, (b) $y$, and (c) $z$ directions.}
\label{classical}
\end{figure}

We also derive the CFI for the BHD, given by 
\begin{equation}\begin{aligned}
 F_C^{\rm (BHD)}=|A_W|^2|\langle f|i\rangle|^2N \text{diag}\left(\sigma^{-2},\sigma^{-2},\frac{1}{k^2}\sigma^{-4} \right).
 \end{aligned}\end{equation}
In Fig.~\ref{classical}, we compare the CFI for the BHD and AD methods, under the same conditions of $\phi=\theta=0.01~\text{rad}$ and $\sigma=0.1$~mm as in Fig.~\ref{quantum}. It shows that the BHD outperforms the AD in measuring the deformation in the $x$ or $y$ direction, as the CFI of the former is more than that of the latter by a factor of $1/\sqrt{\frac{2\sqrt{2/e}}{\pi}} - 1 \approx 35~\%$, while for the deformation in the $z$ direction, the BHD and the AD show the same performance.

\section{Conclusion}

We have presented an experimental scheme for realizing optical real-time detection of the magnetostrictive deformation of a YIG sphere in 3Ds. The deformation induced high-order modes of the scattered field can be extracted by postselection and measured by BHD systems, which allows for the detection of the deformation in 3Ds with the precision up to picometer level. This can be further improved by using a higher-order probe beam. It is worth noting that the second-order HG mode (induced by the waist variation) is for the first time adopted in an optical spatial measurement, which is distinct from other spatial measurements, e.g., the displacement~\cite{1021,034023} and tilt~\cite{173601,0132992,054068} measurement, that are based on the first-order mode induced by the change in the beam axis. Our 3D measurement of the YIG sphere's deformation can be used to characterize the magnomechanical dynamical backaction and the 3D cooling of the mechanical vibration. The real-time detection allows one to conveniently implement feedback control of the mechanical motion to enhance the mechanical cooling or phonon lasing~\cite{121106}. Our method can also be applied to measure the displacement or deformation of nonmagnetic spherical objects.

\clearpage %%%%%%%%% 

\begin{widetext}
\section{Appendix}

\subsection{Evolution of the light field}

Here we provide a detailed derivation of the evolution of the light field (the pointer). The initial input pointer state can be expressed as a quantized field ${\hat{E}_{p}} =i\sqrt{\frac{\hbar\omega}{2\varepsilon_0cT}}\hat{a}_{00}u_{00}$, where $\hat{a}_{00}$ is the annihilation operator of the $\text{HG}_{00}$ mode, $\omega$ is the frequency of the probe beam, $\varepsilon_0$ is the vacuum dielectric constant, and $T$ is the environment temperature. The scattered field can then be obtained
\begin{equation}\begin{aligned} 
\hat{E}_{s}=&Me^{-i\frac{\delta\omega_y}{\omega_y}yp_y-i\frac{\delta\omega_x}{\omega_x}xp_x-i\frac{\delta z}{2k}(p_x^2+p_y^2)} \hat{a}_{00}u_{00},
 \end{aligned}\end{equation}
where $x$ ($p_x$) and $y$ ($p_y$) are the position (momentum) components in the $x$ and $y$ directions, and $\omega_x$ ($\omega_y$) denotes the beam waist radius corresponding to the average deformation displacement in the $x$ ($y$) direction. Here, $M=i\sqrt{\frac{\hbar\omega}{2\varepsilon_0cT}}\frac{\lambda}{2\pi z}\frac{|s_1|+|s_2|}{2}$, with $\lambda$ being the wavelength of the probe beam, and $s_1=\sum_{\iota=1}^\infty \frac{2\iota+1}{\iota(\iota+1)}(a_\iota\pi_\iota+b_\iota\tau_\iota)$ and $s_2=\sum_{\iota=1}^\infty \frac{2\iota+1}{\iota(\iota+1)}(a_\iota\tau_\iota+b_\iota\pi_\iota)$ denote the amplitude functions of the scattered light, where $\pi_\iota=\frac{P_\iota^{(1)}(\cos\Theta)}{\sin\Theta}$, $\tau_\iota=\frac{dP_\iota^{(1)}(\cos\Theta)}{d\Theta}$, with $P_\iota^{(1)}(\cos\Theta)$ being the first-order associated Legendre function and $\Theta$ being the scattering angle. The Mie scattering coefficients are given by $a_\iota=\frac{\Phi_\iota(\alpha)\Phi'_\iota(\varpi\alpha)-\varpi\Phi'_\iota(\alpha)\Phi_\iota(\varpi\alpha)}{\Psi_\iota(\alpha)\Phi'_\iota(\varpi\alpha)-\varpi\Psi'_\iota(\alpha)\Phi_\iota(\varpi\alpha)}$, and $b_\iota=\frac{\varpi\Phi_\iota(\alpha)\Phi'_\iota(\varpi\alpha)-\Phi'_\iota(\alpha)\Phi_\iota(\varpi\alpha)}{\varpi\Psi_\iota(\alpha)\Phi'_\iota(\varpi\alpha)-\Psi'_\iota(\alpha)\Phi_\iota(\varpi\alpha)}$, where $\varpi$ is the relative refractive index of the YIG sphere, $\Phi_\iota(\alpha)=\sqrt{\frac{\pi \alpha}{2}}J_{\iota+\frac{1}{2}}(\alpha)$, $\Psi_\iota(\alpha)=\sqrt{\frac{\pi \alpha}{2}}H_{\iota+\frac{1}{2}}^{(2)}(\alpha)$, with $J_{\iota+\frac{1}{2}}(\alpha)$ and $H_{\iota+\frac{1}{2}}^{(2)}(\alpha)$ being the Bessel function and the second-kind Hankel function of half-integer order, respectively, and the size parameter $\alpha=\frac{2\pi r}{\lambda}$ ($r$ is the radius of the sphere).

The light fields at the dark port 1, bright port 1, dark port 2, and bright port 2 are labeled as $\hat{E}_{d}^1$, $\hat{E}_{b}^1$, $\hat{E}_d^2$, and $\hat{E}_b^2$, respectively, which are obtained as follows:
\begin{equation}\label{ddddd}\begin{aligned} 
{\hat{E}_{d}^1}=&Me^{-i\frac{\delta\omega_x}{\omega_x}xp_x-i\frac{\delta z}{2k}(p_x^2+p_y^2)}[(e^{i\frac{\delta\omega_y}{\omega_y}yp_y+i\frac{\theta}{2}}-{e^{-i\frac{\delta\omega_y}{\omega_y}yp_y-i\frac{\theta}{2}}})/{2}] \hat{a}_{00}u_{00}\\     \approx &Me^{-i\frac{\delta\omega_x}{\omega_x}xp_x-i\frac{\delta z}{2k}(p_x^2+p_y^2)} i\sin\frac{\theta}{2}e^{\frac{\delta\omega_y}{\omega_y}yp_y \cot\frac{\theta}{2}} \hat{a}_{00}u_{00}\\      \approx& M i\sin\frac{\theta}{2}e^{\frac{\delta\omega_y}{\omega_y}yp_y \cot\frac{\theta}{2}} \hat{a}_{00}u_{00},\\  {\hat{E}_{b}^1}=&Me^{-i\frac{\delta\omega_x}{\omega_x}xp_x-i\frac{\delta z}{2k}(p_x^2+p_y^2)} [({e^{-i\frac{\delta\omega_y}{\omega_y}yp_y-i\frac{\theta}{2}}+e^{i\frac{\delta\omega_y}{\omega_y}yp_y+i\frac{\theta}{2}}})/{2}] \hat{a}_{00}u_{00}\\         \approx&Me^{-i\frac{\delta\omega_x}{\omega_x}xp_x-i\frac{\delta z}{2k}(p_x^2+p_y^2)} \cos\frac{\theta}{2}e^{\frac{\delta\omega_y}{\omega_y}yp_y \tan\frac{\theta}{2}} \hat{a}_{00}u_{00}\\         \approx&M\cos\frac{\theta}{2} e^{-i\frac{\delta\omega_x}{\omega_x}xp_x-i\frac{\delta z}{2k}(p_x^2+p_y^2)} \hat{a}_{00}u_{00},\\{\hat{E}_{d}^2}=&M\cos\frac{\theta}{2} e^{-i\frac{\delta z}{2k}(p_x^2+p_y^2)} [(e^{i\frac{\delta\omega_x}{\omega_x}xp_x+i\frac{\phi}{2}}-{e^{-i\frac{\delta\omega_x}{\omega_x}xp_x-i\frac{\phi}{2}}})/{2}] \hat{a}_{00}u_{00} \\         \approx&M i\sin\frac{\phi}{2}\cos\frac{\theta}{2}e^{\frac{\delta\omega_x}{\omega_x}xp_x \cot\frac{\phi}{2}} \hat{a}_{00}u_{00},\\      {\hat{E}_{b}^2}=&M\cos\frac{\theta}{2} e^{-i\frac{\delta z}{2k}(p_x^2+p_y^2)} [({e^{-i\frac{\delta\omega_x}{\omega_x}xp_x-i\frac{\phi}{2}}+e^{i\frac{\delta\omega_x}{\omega_x}xp_x+i\frac{\phi}{2}}})/{2}] \hat{a}_{00}u_{00} \\        \approx&M\cos\frac{\theta}{2}\cos\frac{\phi}{2} e^{-i\frac{\delta z}{2k}(p_x^2+p_y^2)} \hat{a}_{00}u_{00}.
 \end{aligned}\end{equation}
For taking the "$\approx$" sign, we have used the fact that $\frac{\delta\omega_x}{\omega_x},\frac{\delta\omega_y}{\omega_y},\frac{\delta z}{2k}$ are very small quantities. Equation~\eqref{ddddd} agrees with the results obtained in the approach using the quantum formulation, i.e., Eq.~\eqref{bbbbb} in the main text. 

To further determine the weight coefficients of the induced high-order modes, the scattered field can be derived in the following form:
\begin{equation}\begin{aligned} 
\hat{E}_{s}=M &\Bigg(\sqrt{N} \Big(u_{00} +\frac{\delta \omega_x}{\sqrt{2}\omega_x}u_{20} +\frac{\delta \omega_y}{\sqrt{2}\omega_y}u_{02}-i\frac{5\delta z}{4z_R}u_{00}+i\frac{\delta z}{2\sqrt{2}z_R}u_{20}+i\frac{\delta z}{2\sqrt{2}z_R}u_{02} \Big)\\&\,\,\,\, +\frac{\delta \hat{X}_{00}+i\delta \hat{Y}_{00}}{2}u_{00}+\frac{\delta \hat{X}_{20}+i\delta \hat{Y}_{20}}{2}u_{20}+\frac{\delta \hat{X}_{02}+i\delta \hat{Y}_{02}}{2}u_{02} \Bigg) \\  \approx M&\Bigg( \sqrt{N} \Big(u_{00} +\frac{\delta \omega_x}{\sqrt{2}\omega_x}u_{20} +\frac{\delta \omega_y}{\sqrt{2}\omega_y}u_{02}-i\frac{5\delta z}{4z_R}u_{00} \Big)\\&\,\,\,\, +\frac{\delta \hat{X}_{00}+i\delta \hat{Y}_{00}}{2}u_{00}+\frac{\delta \hat{X}_{20}+i\delta \hat{Y}_{20}}{2}u_{20}+\frac{\delta \hat{X}_{02}+i\delta \hat{Y}_{02}}{2}u_{02} \Bigg),
\end{aligned}\end{equation}
where we have introduced the amplitude and phase quadrature operators $\hat{X}_{nm}$ and $\hat{Y}_{nm}$ and the corresponding fluctuation operators $\delta \hat{X}_{nm}$ and $\delta \hat{Y}_{nm}$, e.g., $\hat{X}_{00}$ and $\hat{Y}_{00}$ for the $\text{HG}_{00}$ mode, defined by $\hat{a}_{00}=\frac{\hat{X}_{00}+i\hat{Y}_{00}}{2}$.
It shows that the coefficients of the induced $u_{20}$, $u_{02}$, $iu_{00}$, $iu_{20}$ and $iu_{02}$ modes are $\frac{\epsilon_x}{\sqrt{2}}$, $\frac{\epsilon_y}{\sqrt{2}}$, $\frac{5\delta z}{4z_R}$, $\frac{\delta z}{2\sqrt{2}z_R}$ and $\frac{\delta z}{2\sqrt{2}z_R}$, which are calculated from $\int u_{20}^*u_{00}(\frac{x}{\epsilon_x+1})dxdy$, $\int u_{02}^*u_{00}(\frac{y}{\epsilon_y+1})dxdy$, $\int u_{00}^*u_{00}(\delta z)dxdy$, $\int u_{20}^*u_{00}(\delta z)dxdy$ and $\int u_{02}^*u_{00}(\delta z)dxdy$, respectively, where $\epsilon_x=\frac{\delta\omega_x}{\omega_x}$ and $\epsilon_y=\frac{\delta\omega_y}{\omega_y}$. 
Therefore, the optical fields at the dark port 1, dark port 2, and bright port 2 are given by
\begin{equation} \begin{aligned} 
\hat{E}_d^1/M &=\sqrt{N}i\sin\frac{\theta}{2}\frac{\delta \omega_y \cot\frac{\theta}{2}}{\sqrt{2}\omega_y}u_{02}+\frac{\delta \hat{X}_{02}+i\delta \hat{Y}_{02}}{2}u_{02}, \\          \hat{E}_d^2/M &=\sqrt{N}i\sin\frac{\phi}{2}\cos\frac{\theta}{2} \frac{\delta \omega_x \cot\frac{\phi}{2}}{\sqrt{2}\omega_x}u_{20}+\frac{\delta \hat{X}_{20}+i\delta \hat{Y}_{20}}{2}u_{20}, \\      \hat{E}_b^2/M &=\sqrt{N}\cos\frac{\theta}{2} \cos\frac{\phi}{2} i\frac{5\delta z}{4z_R}u_{00}+\frac{\delta \hat{X}_{00}+i\delta \hat{Y}_{00}}{2}u_{00}.
 \end{aligned}\end{equation} 
Consequently, the corresponding minimum measurable deformations (MMDs) in the $x$, $y$, and $z$ directions are determined as: $\delta\omega_{y{\rm min}}=\frac{\omega_y}{\sqrt{2N}\cos\frac{\theta}{2}}$, $\delta\omega_{x{\rm min}}=\frac{\omega_x}{\sqrt{2N}\cos\frac{\theta}{2} \cos\frac{\phi}{2}}$, $\delta z_{\rm min}=\frac{2z_R}{5\sqrt{N}\cos\frac{\theta}{2} \cos\frac{\phi}{2}}$, which are equivalent to the results presented in the main text.

\subsection{Verification of the induced second-order fields}

The scattering induced second-order fields can be verified via the COMSOL simulation, which are shown in Fig.~\ref{mies}(a), while the far-field interference patterns between the probe and the scattered fields are illustrated in Figs.~\ref{mies}(b) and~\ref{mies}(c).

\begin{figure}[b]
\centering\includegraphics[width=13cm]{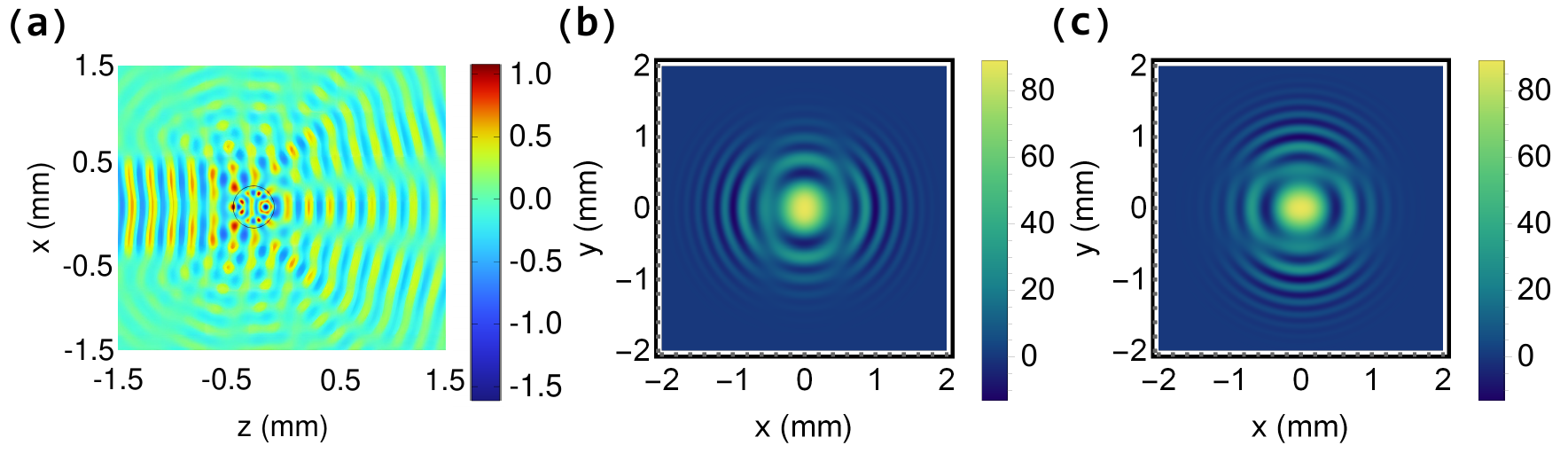}
\caption{(a) The induced second-order fields via the COMSOL simulation. (b)-(c) The far-field interference between the probe and the scattered field for the linear $x$($y$)-polarized probe beam in (b)[(c)]. In (b) and (c), the absorptive loss in the sample is negligible and the poor interference fringe visibility caused by the intensity difference between the probe and scattered field is improved. We take $\lambda = 125~\mu m$, $\omega_0=150~\mu m$, the refractive index $\varpi = 2.19$, the radius of the YIG sphere $r = 125$~$\mu$m, and the focal length of the objective lens is 0.002~m.}
\label{mies}
\end{figure}

\subsection{Derivation of the interaction Hamiltonian}
 
In this section, we derive in detail the general translation operators corresponding to the deformations $\delta\omega_x$, $\delta\omega_y$ and $\delta z$, respectively, from which we can obtain the interaction Hamiltonian. For the probe beam being the $\text{HG}_{nm}$ mode, when the waist size undergoes a small variation $\delta\omega_x$ in the $x$ direction, its transverse distribution at the beam waist can be expressed as
$\psi_{nm}^{\epsilon_x}=\frac{1}{\sqrt{\epsilon_x+1}}\psi_{nm}(\frac{x}{\epsilon_x+1},y,0)$,
where $\epsilon_x=\frac{\delta\omega_x}{\omega_x}$.
The $\text{HG}_{nm}$ mode beam is the solution of the equation 
\begin{equation}\begin{aligned}
&\frac{\partial}{\partial \epsilon_x}\psi_{nm}^{\epsilon_x}=-\Big(\frac{1}{2}+x\frac{\partial}{\partial x}\Big)\psi_{nm}^{\epsilon_x},
\end{aligned}\end{equation}
which can be rewritten as
\begin{equation}\begin{aligned}
\frac{d}{d\epsilon_x}|\psi_{nm}^{\epsilon_x}\rangle &= -\frac{i}{2}(\hat{X}\hat{P}_x + \hat{P}_x\hat{X})|\psi_{nm}^{\epsilon_x}\rangle
\end{aligned}\end{equation}
assuming $\hbar=1$. The above equation has the formal solution
\begin{equation}\begin{aligned}|\psi_{nm}^{\epsilon_x}\rangle&=\hat{U}({\epsilon_x})|\psi_{nm}\rangle,\end{aligned}\end{equation}
where the corresponding evolution operator $\hat{U}({\epsilon_x})=e^{-i\frac{\hat{X}\hat{P}_x + \hat{P}_x\hat{X}}{2}\epsilon_x}$.

When the waist size has a small variation $\delta\omega_y$ in the $y$ direction, its transverse distribution at the beam waist is given by
$\psi_{nm}^{\epsilon_y}=\frac{1}{\sqrt{\epsilon_y+1}}\psi_{nm}(x,\frac{y}{\epsilon_y+1},0)$,
with $\epsilon_y=\frac{\delta\omega_y}{\omega_y}$. In the same way, we can obtain the corresponding evolution operator, i.e., $\hat{U}({\epsilon_y})=e^{-i\frac{\hat{Y}\hat{P}_y + \hat{P}_y\hat{Y}}{2}\epsilon_y}$.

When the waist position of the $\text{HG}_{nm}$ beam undergoes a small shift $\delta z$ in the propagating $z$ direction, its transverse distribution is obtained as

$\psi_{nm}^{\delta z}=\sqrt{\frac{\omega_x}{\omega_z}}\sqrt{\frac{\omega_y}{\omega_z}}\psi_{nm}(\frac{\omega_x}{\omega_z}x,\frac{\omega_y}{\omega_z}y,\delta z)e^{\frac{ikx^2}{2R_x(\delta z)}}e^{\frac{iky^2}{2R_y(\delta z)}}e^{-i(n+1/2)\arctan\frac{\delta z}{z_{Rx}}}e^{-i(m+1/2)\arctan\frac{\delta z}{z_{Ry}}}$, where $R_x(\delta z)$ and $R_y(\delta z)$ denote the curvature radii in the $x$ and $y$ directions at $\delta z$, respectively. 

The $\text{HG}_{nm}$ beam is the solution of the equation
\begin{equation}
\begin{aligned}
&-i\frac{\partial}{\partial z}\psi_{nm}^{\delta z}=\frac{1}{2k}\nabla^2\psi_{nm}^{\delta z}.
\end{aligned}
\end{equation}
which can be rewritten as
\begin{equation}
\begin{aligned}
\frac{d}{dz}|\psi_{nm}^{\delta z}\rangle&=-\frac{i}{2k}\hat{P}^2|\psi_{nm}^{\delta z}\rangle.
\end{aligned}
\end{equation}
The formal solution of the above equation is
\begin{equation}
\begin{aligned}
|\psi_{nm}^{\delta z}\rangle&=\hat{U}(\delta z)|\psi_{nm}\rangle, 
\end{aligned}
\end{equation}
where the evolution operator $\hat{U}(\delta z)=e^{-\frac{i}{2k}\hat{P}^2\delta z}$. 

In summary, the whole evolution operator can be expressed as $\hat{U}=\hat{U}(\epsilon_x)\hat{U}(\epsilon_y)\hat{U}(\delta z)$ for $t\geq t_0$, and $\hat{U}=0$ for $t < t_0$, where $t_0$ is the time when the scattering begins. The total interaction Hamiltonian can be obtained by combining the postselection system and the interaction during the scattering process ($i\frac{\partial\ln{\hat{U}}}{\partial t}$), given by:
\begin{equation}
\hat{H}_{\rm int}=\Big(\frac{\delta\omega_y}{\omega_y}\hat{A}_1\otimes\frac{\hat{Y}\hat{P}_y + \hat{P}_y\hat{Y}}{2}+\frac{\delta\omega_x}{\omega_x}\hat{A}_2\otimes\frac{\hat{X}\hat{P}_x + \hat{P}_x\hat{X}}{2} +\frac{\delta z}{2k}\hat{A}_3\otimes\hat{P}^2\Big)\delta(t-t_0),
\end{equation}
which is Eq.~\eqref{HHHH} in the main text.

This can also be understood from the transformations in the phase space~\cite{Wigner1980}. The transformations ${(\hat{X}\hat{P}_x + \hat{P}_x\hat{X})}/{2}$ and ${(\hat{Y}\hat{P}_y + \hat{P}_y\hat{Y})}/{2}$ correspond to the scaling operations, which modify the wave packet distribution while preserving its volume. Specifically, under these operations, the Wigner function makes the transformations $W(x,p_x)\xrightarrow{{(\hat{X}\hat{P}_x + \hat{P}_x\hat{X})}/{2}} W(\frac{x}{1+\epsilon_x},p_x(1+\epsilon_x))$ and $W(y,p_y)\xrightarrow{({\hat{Y}\hat{P}_y + \hat{P}_y\hat{Y})}/{2}} W(\frac{y}{1+\epsilon_y},p_y(1+\epsilon_y))$, with the scaling factor $1+\epsilon_x$ and $1+\epsilon_y$. The transformation $\hat{P}^2/(2k)$ corresponds to the shearing transformation, which stretches the spatial distribution while maintaining the momentum distribution, thereby yielding an increase in $\omega(z)$. The corresponding transformation of the Wigner function is $W(q,p)\xrightarrow{\hat{P}^2/(2k)} W(q+\frac{p}{k}\delta z,p)$.

\subsection{Final pointer state for a general \texorpdfstring{$\text{HG}$}{HG} probe beam}

In this section, we derive the final state in our scheme for the probe beam being a general $\text{HG}_{nm}$ mode. According to the properties of Hermite polynomials, the following recurrence relation for the pointer $\psi_{nm}$ can be obtained
\begin{equation}\begin{aligned}
\partial_x\psi_{nm} &=\frac{1}{\omega_x} \left(\sqrt{n}\psi_{n-1,m}-\sqrt{(n+1)}\psi_{n+1,m} \right) , \\ 
x\psi_{nm} &=\frac{\omega_x}{2} \left(\sqrt{n}\psi_{n-1,m}+\sqrt{(n+1)}\psi_{n+1,m} \right), \\
\partial_y\psi_{nm} &=\frac{1}{\omega_y} \left(\sqrt{m}\psi_{n,m-1}-\sqrt{(m+1)}\psi_{n,m+1} \right), \\ 
y\psi_{nm} &=\frac{\omega_y}{2} \left(\sqrt{m}\psi_{n,m-1}+\sqrt{(m+1)}\psi_{n,m+1} \right).
\end{aligned}\end{equation}
Using the above relations, the final state $|\psi_f\rangle$ in Eq.~\eqref{psi} can be further expressed as
\begin{equation}\label{psiff}
\begin{aligned}
 |\psi_f\rangle=&\sum_{j=1}^3\langle f_j|i_j\rangle|\psi_{nm}\rangle-\langle f_3|i_3\rangle\left(\frac{iA_{W_3}}{2k\omega_{x}^2}\delta z(2n+1)+\frac{iA_{W_3}}{2k\omega_{y}^2}\delta z(2m+1) \right)|\psi_{nm}\rangle\\&-\langle f_3|i_3\rangle\left(\frac{iA_{W_3}}{k\omega_{x}^2}\delta z(2n+1)+\frac{iA_{W_3}}{k\omega_{y}^2}\delta z(2m+1) \right)|\psi_{nm}\rangle\\&+\left(\langle f_3|i_3\rangle\frac{iA_{W_3}}{2k\omega_{x}^2}\delta z-\langle f_2|i_2\rangle\frac{A_{W_2}}{2\omega_x}\delta\omega_x \right)\sqrt{n(n-1)}|\psi_{n-2,m}\rangle\\&+\left(\langle f_3|i_3\rangle\frac{iA_{W_3}}{2k\omega_{y}^2}\delta z-\langle f_1|i_1\rangle\frac{A_{W_1}}{2\omega_y}\delta\omega_y \right)\sqrt{m(m-1)}|\psi_{n,m-2}\rangle\\&+ \left(\langle f_2|i_2\rangle\frac{A_{W_2}}{2\omega_x}\delta\omega_x+\langle f_3|i_3\rangle\frac{iA_{W_3}}{2k\omega_{x}^2}\delta z \right)\sqrt{(n+1)(n+2)}|\psi_{n+2,m}\rangle\\& +\left(\langle f_1|i_1\rangle\frac{A_{W_1}}{2\omega_y}\delta\omega_y+\langle f_3|i_3\rangle\frac{iA_{W_3}}{2k\omega_{y}^2}\delta z \right)\sqrt{(m+1)(m+2)}|\psi_{n,m+2}\rangle\\
\approx &\sum_{j=1}^3\langle f_j|i_j\rangle|\psi_{nm}\rangle-\langle f_3|i_3\rangle \left(\frac{iA_{W_3}}{2k\omega_{x}^2}\delta z(2n+1)+\frac{iA_{W_3}}{2k\omega_{y}^2}\delta z(2m+1) \right)|\psi_{nm}\rangle\\&-\langle f_3|i_3\rangle\left(\frac{iA_{W_3}}{k\omega_{x}^2}\delta z(2n+1)+\frac{iA_{W_3}}{k\omega_{y}^2}\delta z(2m+1) \right)|\psi_{nm}\rangle\\&+\langle f_2|i_2\rangle\left(\frac{A_{W_2}}{2\omega_x}\delta\omega_x\sqrt{(n+1)(n+2)}|\psi_{n+2,m}\rangle-\frac{A_{W_2}}{2\omega_x}\delta\omega_x\sqrt{n(n-1)}|\psi_{n-2,m}\rangle\right)\\&+\langle f_1|i_1\rangle \left(\frac{A_{W_1}}{2\omega_y}\delta\omega_y\sqrt{(m+1)(m+2)}|\psi_{n,m+2}\rangle-\frac{A_{W_1}}{2\omega_y}\delta\omega_y\sqrt{m(m-1)}|\psi_{n,m-2}\rangle\right)\\
=&\left(i\sin{\frac{\theta}{2}}+i\sin{\frac{\phi}{2}}\cos{\frac{\theta}{2}}+\cos{\frac{\theta}{2}}\cos{\frac{\phi}{2}}\right)|\psi_{nm}\rangle\\&
-\cos{\frac{\theta}{2}}\cos{\frac{\phi}{2}}\left(\left(\frac{i\delta z}{2k\omega_{x}^2}+\frac{i\delta z}{k\omega_{x}^2}\right)(2n+1)+\left(\frac{i\delta z}{2k\omega_{y}^2}+\frac{i\delta z}{k\omega_{y}^2}\right)(2m+1) \right)|\psi_{nm}\rangle\\&
+i\sin{\frac{\phi}{2}}\cos{\frac{\theta}{2}}\left(\frac{\delta\omega_x\cot{\frac{\phi}{2}}}{2\omega_x}\sqrt{(n+1)(n+2)}|\psi_{n+2,m}\rangle-\frac{\delta\omega_x\cot{\frac{\phi}{2}}}{2\omega_x}\sqrt{n(n-1)}|\psi_{n-2,m}\rangle\right)\\&+i\sin{\frac{\theta}{2} }
\left(\frac{\delta\omega_y\cot{\frac{\theta}{2}}}{2\omega_y}\sqrt{(m+1)(m+2)}|\psi_{n,m+2}\rangle-\frac{\delta\omega_y\cot{\frac{\theta}{2}}}{2\omega_y}\sqrt{m(m-1)}|\psi_{n,m-2}\rangle\right).
\end{aligned}
\end{equation}
It tells that $|\psi_f\rangle$ contains the components ($|\psi_{n-2,m}\rangle$, $|\psi_{n+2,m}\rangle$), ($|\psi_{n,m-2}\rangle$, $|\psi_{n,m+2}\rangle$) and $i|\psi_{n,m}\rangle$, which correspond to the deformations $\delta\omega_x$, $\delta\omega_y$, and $\delta z$, respectively. Similarly,  $\frac{\delta z(2n+1)}{k\omega_{x}^2}+\frac{\delta z(2m+1)}{k\omega_{y}^2}$ is the deformation-induced Gouy phase shift. Note that, here $i|\psi_{n-2,m}\rangle$, $i|\psi_{n,m-2}\rangle$, $i|\psi_{n+2,m}\rangle$ and $i|\psi_{n,m+2}\rangle$ are neglected due to their small proportions. For a fundamental mode pointer, i.e., $n=m=0$, Eq.~\eqref{psiff} reduces to Eq.~\eqref{psif}.
\end{widetext}

%%%%%%%%% References %%%%%%%%%%%%
\bibliography{doc/latex/revtex/aps/ref}
\end{document}